\documentclass{tex-library/jfm}

\usepackage{graphicx, amsmath, amssymb, amsfonts, mathtools, mathrsfs, color}
\usepackage{natbib, comment, todonotes}
\usepackage[justification=RaggedRight]{caption}
\usepackage{xcolor}
\usepackage{soul}
\usepackage[colorlinks]{hyperref}
\usepackage[capitalise]{cleveref}
\usepackage{bm}
\usepackage[font=small,justification=justified]{caption}
\usepackage{subcaption}
\usepackage[english]{babel}
\usepackage[toc,page]{appendix}
\usepackage{url}
\usepackage[normalem]{ulem}
\usepackage{mathbbol}
\usepackage{upgreek}
\usepackage{tikz}
\usetikzlibrary{tikzmark}

\newcommand{\pd}[2]{ \frac{ \partial #1}{ \partial #2 } }

\newcommand{\Lap}{\grad^2}

\newcommand{\bvec}[1]{\ensuremath{\boldsymbol{#1}}}
\newcommand{\grad}{\nabla}

\DeclarePairedDelimiter\floor{\lfloor}{\rfloor}

\newcommand{\uu}{\bvec{u}}

\newcommand{\T}{\theta}
\newcommand{\Pra}{\text{Pr}}
\newcommand{\Ra}{\text{Ra}}
\newcommand{\ReN}{\text{Re}}

\newcommand{\Nu}{\text{Nu}}
\newcommand{\RB}{Rayleigh-B\'enard}
\newcommand{\RBC}{\RB{} convection}

\newcommand{\upii}{u_p^{(i)}}

\newcommand{\second}{$2^\textnormal{nd}$}

\newcommand{\updel}{\Updelta}
\newcommand{\im}{\mathbb{i}}
\newcommand{\Cr}{\mbox{\textit{Cr}}}
\newcommand{\mass}{m}

\shorttitle{Covering convection with thermal blankets}
\shortauthor{J. M. Huang}

\title{Covering convection with thermal blankets: fluid-structure interactions in thermal convection}

\author{Jinzi Mac Huang\aff{1,2}\corresp{\email{machuang@nyu.edu}}}

\affiliation{
\aff{1}NYU-ECNU Institute of Physics and Institute of Mathematical Sciences, New York University Shanghai, Shanghai, China
\aff{2}Applied Math Lab, Courant Institute, New York University, New York, USA
}

\begin{document}

\maketitle
\begin{abstract}
The continental plates of Earth are known to drift over a geophysical timescale, and their interactions have lead to some of the most spectacular geoformations of our planet while also causing natural disasters such as earthquakes and volcanic activity. Understanding the dynamics of interacting continental plates is thus significant. In this work, we present a fluid mechanical investigation of the plate motion, interaction, and dynamics. Through numerical experiments, we examine the coupling between a convective fluid and plates floating on top of it. With physical modeling, we show the coupling is both mechanical and thermal, leading to the thermal blanket effect: the floating plate is not only transported by the fluid flow beneath, it also prevents the heat from leaving the fluid, leading to a convective flow that further affects the plate motion. By adding several plates to such a coupled fluid-structure interaction, we also investigate how floating plates interact with each other and show that, under proper conditions, small plates can converge into a supercontinent. 
\end{abstract}

\begin{keywords}
Authors should not enter keywords on the manuscript, as these must be chosen by the author during the online submission process and will then be added during the typesetting process (see http://journals.cambridge.org/data/\linebreak[3]relatedlink/jfm-\linebreak[3]keywords.pdf for the full list)
\end{keywords}

\section{Introduction}
\label{intro}

\begin{figure}
\centering
\includegraphics[width=0.85\textwidth]{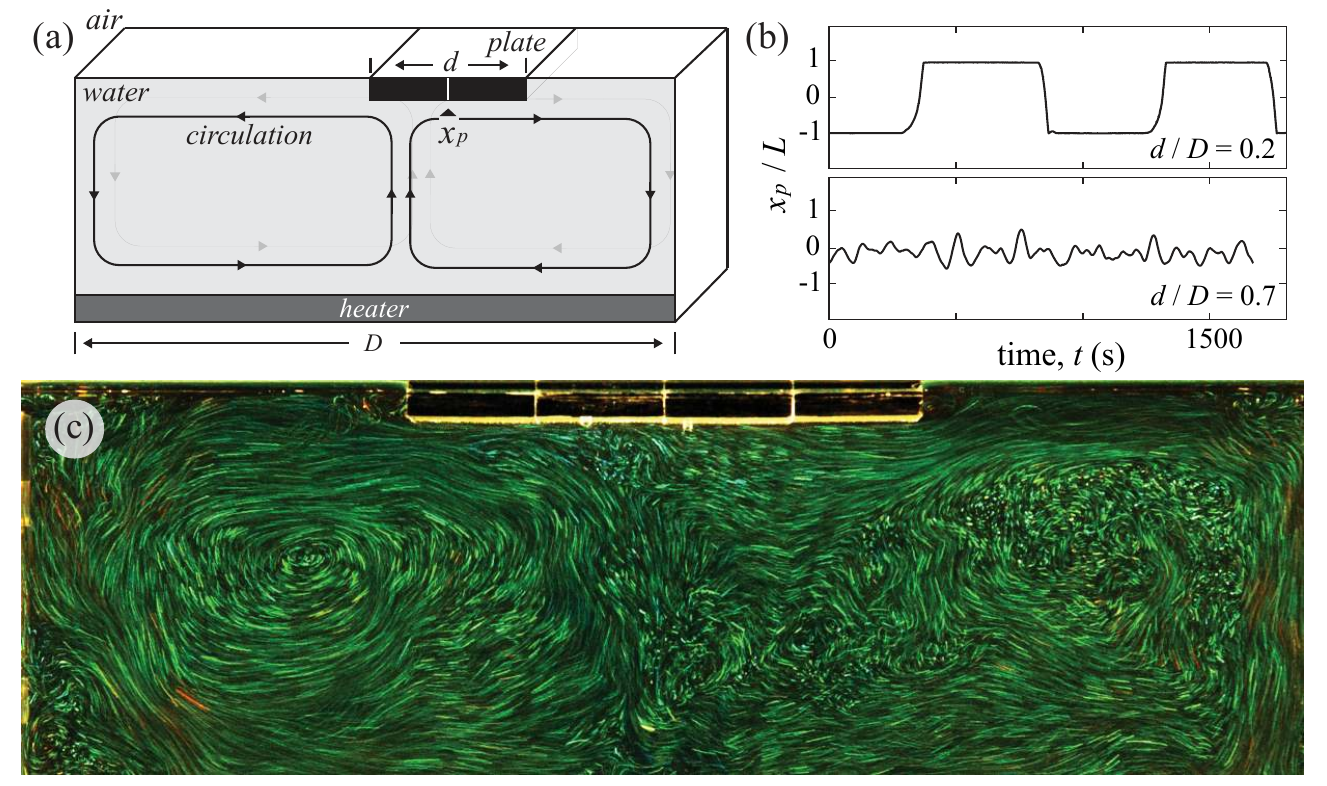}
\caption{ \RBC{} coupled to a free-floating plate leads to different dynamics of plate motion. (a) Schematics of the interaction between \RBC{} and the floating plate. The fluid is heated from below and has an open free surface, the floating plate on this free surface is transported by the fluid force exerted from below. (b) Different states of plate motion. A small plate with $d/D=0.2$ oscillates between two sidewalls of the convection cell, while a big plate with $d/D = 0.7$ is trapped in the middle of the convection cell. Here $L = (D-d)/2$ is the bound of plate center $x_p$. (c) Flow visualization reveals two counter-rotating large-scale circulations when the plate is located at the center of the convection cell. Image credit: \cite{zhong2007b,mac2018stochastic}. }
\label{fig1}
\end{figure}

Fluid-structure interactions appear at many different scales on our planet, and perhaps the largest one is the continental plate tectonics \citep{plummer2001physical}. It is believed that this tectonic motion results from the thermal convection in Earth's mantle \citep{kious1996dynamic}, where the mantle materials behave like a fluid \citep{turcotte2002geodynamics} that is heated from the core and cooled at the surface of Earth. Under gravity, this configuration of heating and cooling leads to thermal convection, whose circulation provides forcing to the continental plates through shearing. This is considered to be the simplest picture of plate tectonics, however many details of the plate dynamics, like the periodic formation of supercontinents and the associated geological Wilson cycle \citep{Burke2011}, require further investigation.

Laboratory experiments have proven to be an effective way of understanding the fluid-structure interaction behind the plate tectonics, with many successful studies that couple the thermally convective fluid to solid structures \citep{elder1967convective,howard1970self,Whitehead1972}. Aimed at recovering the plate dynamics and understanding the associated fluid-structure interactions, a series of laboratory experiments was later conducted by \cite{zhang2000periodic,Zhong2005,zhong2007a,zhong2007b}. Shown in \cref{fig1}(a), this experiment employs water as the working fluid, and a heater beneath provides heating while the ventilation at water-air interface provides cooling, resulting in Rayleigh-B\'enard convection \citep{Ahlers2009}. A floating plate of size $d$ is carefully placed on top of this convective domain of total length $D$. This moving plate with center location $x_p$ has only fluid force acting on it, unless it hits the wall on the left or the right. Large-scale circulations \citep{Araujo2005,Brown2007,moore2023fluid} are observed to form in the convective fluid, as shown in \cref{fig1}(a) and \cref{fig1}(c). Depending on the location $x_p$, the plate can be either transported by the circulating fluid, or situated on top of a converging or a diverging center of the surface flow [this is the case shown in \cref{fig1}(a)]. Interestingly, different plate motions are observed depending on its size: When the ratio between plate size $d$ and tank width $D$ is smaller than a critical number near $0.58$, the plate oscillates between the two sidewalls as shown in \cref{fig1}(b); When this ratio is above the critical value, the plate is trapped at the center of tank, as shown in \cref{fig1}(b) and \cref{fig1}(c).

Zhong, Zhang, and others investigated this transition of behaviors, and they discovered that the so-called thermal blanket effect is responsible here \citep{zhang2000periodic,Zhong2005,zhong2007a,zhong2007b,mac2018stochastic,mao2019dynamics,mao2021insulating,PhysRevE.109.025102}. In this theory, the floating plate serves as an insulator (like a blanket) on top of the convective fluid, hence the fluid beneath becomes warmer due to the lack of ventilation. The warm fluid then rises, creating a diverging surface flow as shown in \cref{fig1}(a) that can transport the plate. The coupling between the fluid and the floating plate therefore goes two ways: the plate modifies the flow temperature and leads to thermal convection; the formed convective flows transport the floating plate. Their interplay leads to nontrivial dynamics of the plate shown in \cref{fig1}(b), and the physically inspired Zhong-Zhang model \citep{Zhong2005,zhong2007a,zhong2007b} successfully captures the transition of dynamics. Recently, more careful investigations on the Zhong-Zhang model have lead to new advancements in the stochastic \citep{mac2018stochastic} and dynamical \citep{PhysRevE.109.025102} modeling of fluid-structure interactions.

While the laboratory experiments are conducted in a domain of fluid with finite size, numerical simulations can be conducted in a domain that resembles the mantle of Earth. The numerical work of \cite{gurnis1988large} provides one of the first time-dependent simulations of continental drift, where the fluid domain is 2-dimensional and periodic. After this, many more numerical works have investigated the details of continental drift \citep{zhong1993dynamic,lowman1993mantle,lowman1995mantle,lowman1999thermal,lowman1999effects,lowman2001influence,zhong2000role}, engaging higher resolutions, more detailed modeling of fluid-structure interactions, and 3-dimensional simulations of the interior of Earth. In recent years, the mobility of the continental plate has become a focus of numerical study, where persistent motion is observed for larger plates while small plates tend to move sporadically \citep{gurnis1988large,Whitehead2015,mao2019dynamics,mao2021insulating}. In these works, the thermal blanket effect is once again recognized as an important factor causing the diverse plate dynamics. 

This work is a continuation of an earlier investigation, \cite{Huang_2024}, where the thermal and mechanical coupling between one floating plate and convective fluid is modelled through a simple stochastic model. This model shows that the covering ratio $\Cr$, defined as the ratio of the plate area to the total surface area, is a direct measure of the thermal blanket effect. A critical covering ratio $\Cr^*$ is identified to distinguish the dynamics of the plate: For a small plate with $\Cr\ll\Cr^*$, the plate is passive to the flow field and exhibits little motion; For a plate with $\Cr \gg \Cr^*$, the strong thermal blanket effect leads to persistent plate translation. For plates with $\Cr\approx\Cr^*$, more complicated transitioning dynamics are observed.

In this work, we first introduce an efficient 2-dimensional spectral solver that can accurately capture the motions and interactions of multiple floating plates on top of \RBC{}. In a periodic domain shown in \cref{fig2}, this solver can handle the Navier-Stokes flows presented in laboratory experiments, with simple modifications available for the geophysical Stokes flows in the mantle. Moreover, multiple floating plates can be simulated as fast as a single plate problem, as the floating plates are simply treated as an area with different boundary conditions. A specially-tailored spectral solver handles the resulting inhomogeneous Robin conditions for both the temperature and the stream function, allowing for efficient time-stepping and spectral accuracy. This enables us to systematically introduce 1, 2, and many floating plates, and to show how the thermal blanket effect dictates their interactions with the convective flow beneath and each other. The covering ratio $\Cr$ is once again identified as a key factor affecting the plate dynamics and the stable formation of supercontinents.

This paper is arranged as follows: In \cref{math}, we will mathematically formulate the \RBC{} and its coupling to the plate motion; In \cref{numericalmethods}, a numerical scheme and its implementation for solving this free-boundary problem will be introduced; In \cref{results}, numerical simulations of single, double, and multiple plate dynamics will be included and discussed; Finally, we will discuss extensions and future works in \cref{discussion}.

\section{Mathematical formulation}
\label{math}

\subsection{Flow and temperature equations}
We consider a dimensionless set of equations by rescaling the length scale by the cell height $H$, the time scale by the diffusion time $H^2/\kappa$ ($\kappa$ is thermal diffusivity), and temperature by the temperature difference $\updel T$ between the heater and free surface. The $x$ direction of the fluid domain is periodic with period $\Gamma = D/H$ ($D$ is the domain width), so the overall computational domain is $x\in(0,\Gamma)$, $y\in(0,1)$ as shown in \cref{fig2}. With the Boussinesq approximation, the resulting PDEs for flow speed $\uu = (u,v)$, pressure $p$, and temperature $\T \in [0,1]$ are
\begin{align}
    &\frac{D\uu}{Dt} = -\nabla p + \Pra{}\,\Lap\uu + \Ra{}\,\Pra{}\, \T,\label{NS}\\
    &\nabla\cdot \uu = 0,\\
    &\frac{D \T}{Dt} = \Lap \T.
\end{align}

Two dimensionless numbers appear during this non-dimensionalization: the Rayleigh number $\Ra{} = \alpha g \updel T H^3 /\nu\kappa$ and the Prandtl number $\Pra{} = \nu/\kappa$, with $\nu, \alpha$ and $g$ denoting kinematic viscosity, thermal expansion coefficient of the fluid, and the acceleration due to gravity. We have further assumed that the physical properties of the fluid depend on temperature weakly, so $\Ra{}$ and $\Pra{}$ do not depend on $\theta$.

As our simulation is 2-dimensional, it is convenient to write the Navier-Stokes equation in a vorticity \& steam function format \citep{peyret2002spectral}, 
\begin{align}
\frac{D \omega}{Dt}  &= \Pra{} \Lap \omega   + \Pra{} \Ra{} \frac{\partial \T}{\partial x},\label{omegaeq}\\
 -\Lap \psi &= \omega,\ \  \uu = \nabla_\perp \psi,\label{psieq}\\
 \frac{D \T}{Dt}  &=  \Lap \T. \label{Teq}
\end{align}
Instead of directly solving for $\uu$ and $p$, the $z$-component of vorticity $\omega = \mathbf{k}\cdot\nabla\times \uu$ and the stream function defined by $\uu = \nabla_\perp \psi = (\psi_y, -\psi_x)$ are solved first.

\subsection{Boundary conditions}
\begin{figure}
\centering
\includegraphics[width=0.7\textwidth]{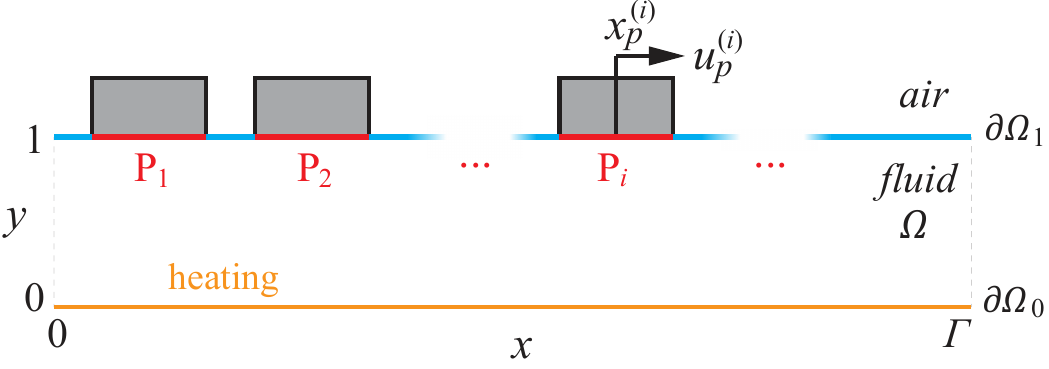}
\caption{Schematics of the floating plate problem. The fluid domain $\Omega$ is heated from the bottom surface $\partial\Omega_0$ and has an open surface on top ($\partial\Omega_1$), floating plates $P_1, P_1, P_2,\dots$ cover part of this open surface and shield the heat from leaving the fluid. }
\label{fig2}
\end{figure}

While \cref{omegaeq,psieq,Teq} are standard for modeling \RBC{}, the boundary conditions become complicated when floating plates are present. Shown in \cref{fig2}, the fluid domain $\Omega$ is bounded between the bottom heating wall $y=0$ ($\partial\Omega_0$) and the top free surface $y=1$ ($\partial\Omega_1$). The segments of top surface covered by the floating plates are labelled as $P_1, P_2, \cdots$, whose centers are $x_p^{(1)}, x_p^{(2)}, \cdots$. 

The boundary conditions for the bottom heating wall are set straightforwardly as constant temperature and no-slip,
\begin{equation}
    \T = 1,\,  u=v=0 \quad \mbox{at  } y = 0.
\end{equation}
For the vorticity-stream function format,
\begin{equation}
\label{bottom_cond}
    \T = 1,\,  \psi=\psi_y=0 \quad \mbox{at  } y = 0.
\end{equation}

The top condition consists of two types of regions: for the free surface (not covered by $P_i$), the temperature is low and the flow is shear free; for the region covered by the plate $P_i$, the heat flux is zero and the flow shares the same velocity as the plate. The zero-flux condition originates from the ``thermal blanket" effect caused by the low heat conduction of solids.  The boundary conditions at $y = 1$ can be summarized as
\begin{align}
    \T = 0,\,  u_y=v=0 \quad &\mbox{for } y=1 \mbox{ and } x\notin \cup P_i,\\
    \T_y = 0,\,  u = u_p^{(i)}, v=0 \quad &\mbox{for } y=1 \mbox{ and } x\in P_i.
\end{align}

Here $u_p^{(i)} = \dot{x}_p^{(i)}$ is the velocity of $i$th plate $P_i$. For convenience, we can also write the top boundary conditions in a more compact way, 
\begin{equation}
\label{movingBC}
  \begin{cases}
    (1-a)\,\T + a\,\T_y = 0\\
    a\,u+(1-a)\,u_y = g\quad\quad\mbox{at } y = 1.\\
    v = 0
\end{cases}  
\end{equation}
For the vorticity-stream function format,
\begin{equation}
\label{movingBC-psi}
  \begin{cases}
    (1-a)\,\T + a\,\T_y = 0\\
    a\,\psi_y+(1-a)\,\psi_{yy} = g\quad\quad\mbox{at } y = 1.\\
    \psi = 0
\end{cases}  
\end{equation}
Here $a(x) = \sum_{i}\mathbb{1}_{x\in P_i}$, $g(x) =  \sum_{i}u_p^{(i)}\mathbb{1}_{x\in P_i}$, and $\mathbb{1}_{x\in P_i}$ is the characteristic function such that 
\begin{equation}
    \mathbb{1}_{x\in P_i} = \begin{cases}
    1 \quad \mbox{ if } x\in P_i,\\
    0 \quad \mbox{ otherwise. }
\end{cases}  
\end{equation}

On plate $P_i$, two types of forces drive its motion: the fluid force $f^{(i)}$ due to the shear from convective flows;  the interacting force $f_l^{(i)}$ or $f_r^{(i)}$ when the left or right neighboring plate ($P_{i-1}$ or $P_{i+1}$) makes contact with plate $P_i$.

For the fluid force, we simply integrate the fluid shear stress, 
\begin{equation}
    f^{(i)} = -\Pra\int_{P_i} u_y(x,1,t)\, dx = -\Pra\int_0^\Gamma u_y(x,1,t)\mathbb{1}_{x\in P_i}d\,x.
\end{equation}

The interaction forces $f_l^{(i)}$ and $f_r^{(i)}$ are modelled as a short-range interaction force to ensure a fully-elastic collision between plates. The numerical implementations will be included in \cref{num-boundary-dynamics}.

Finally, we add all the forces together and evolve the plate location as
\begin{align}
    \dot{x}_p^{(i)} &= \upii{}, \label{xpdot}\\
    \dot{u}_p^{(i)}  &= a_p^{(i)} = \mass^{-1}\left[\,f_l^{(i)}+f_r^{(i)}-\Pra \int_0^{\Gamma} u_y(x,1,t) \mathbb{1}_{x\in P_i}\, dx\right].\label{updot}
\end{align}
Here $a_p^{(i)}$ is the acceleration of $P_i$ and $\mass$ is the dimensionless mass of plate.

\section{Numerical methods}
\label{numericalmethods}

\subsection{Time stepping}
We discretize time with the second order Adam-Bashforth Backward Differentiation method (ABBD2). At time step $t_n = n\updel T$, \cref{omegaeq,Teq,psieq} become
\begin{align}
\label{omegadisc}
\Lap \omega_n -\sigma_1\omega_n  &= f_n,\\
\label{Tdisc}
\Lap \T_n -\sigma_2 \T_n  &= h_n,\\
\label{psidisc}
 -\Lap \psi_n &= \omega_n,
\end{align}
where 
\begin{align}
\sigma_1 &= \frac{3}{2\,\Pra\,\updel t},\quad \sigma_2 = \frac{3}{2\updel t}, \\[8pt]
f_n &= \Pra^{-1}\left[ 2  (\uu \cdot \grad \omega)_{n-1} - (\uu \cdot \grad \omega)_{n-2}\right]  \label{fn}\\
&\hspace{40pt}- (2\, \Pra\, \updel t)^{-1} \left(4\omega_{n-1}-\omega_{n-2}\right) -\Ra\, \left(\pd{\T}{x}\right)_{n},\notag \\[8pt]
h_n &= \left[ 2  (\uu \cdot \grad \T)_{n-1} - (\uu \cdot \grad \T)_{n-2}\right] - (2\updel t)^{-1} \left(4\T_{n-1}-\T_{n-2}\right). \label{hn}
\end{align}

\Cref{omegadisc,Tdisc,psidisc} are Helmholtz equations that can be solved by standard numerical methods \citep{peyret2002spectral}, and this implicit-explicit operator splitting scheme has been used in many numerical studies of thermal convection \citep{Huang2022a,Huang_2024}. However, modifications have to be made to accommodate the inhomogeneous Robin boundary conditions \cref{movingBC-psi}. We will detail this modified Helmholtz solver in Appendices A-C, and a flow chart of the numerical procedure can be found in Appendix D.

In \cref{fn,hn}, nonlinear terms like $\uu \cdot \grad \T$ and $\uu \cdot \grad \omega$ are computed pseudo-spectrally, with a simple and efficient anti-aliasing filter \citep{Hou2007}. With given initial and boundary data, \eqref{Tdisc} can be solved first to obtain $\T_{n}$, which is inserted in $f_{n}$ so \eqref{omegadisc} can be solved next. Finally, \eqref{psidisc} is solved with the known $\omega_{n}$. The spatial and temporal resolution of our study is also detailed in Appendix D.

After solving for the flow and temperature fields, the acceleration $a_{p,\,n}^{(i)}$ of plate $P_i$ can be computed via \cref{updot}, and \cref{xpdot,updot} are integrated with a \second{} order Adam-Bashforth method, 
\begin{align}
x_{p,\,n+1}^{(i)} &=  x_{p,\,n}^{(i)}+\frac{\updel t}{2}\left[ 3u_{p,\,n}^{(i)} - u_{p,\,n-1}^{(i)}\right],\label{loc}\\
u_{p,\,n+1}^{(i)} &=  u_{p,\,n}^{(i)}+\frac{\updel t}{2}\left[ 3a_{p,\,n}^{(i)} - a_{p,\,n-1}^{(i)}\right].\label{spd}
\end{align}

\subsection{Smooth boundary conditions}
\label{num-smooth-bd}
In principle, the introduced fluid and heat solver is able to manage the inhomogeneous Robin boundary condition \cref{movingBC} at $y=1$. However this boundary condition is not smooth, therefore limiting the accuracy of a numerical method with finite resolution. To overcome this, we aim to construct a smooth characteristic function $\hat{\mathbb{1}}_{x\in P_i}$ so it is compactly supported and sufficiently smooth.

We first construct a smooth step function in 1D, whose derivative $\phi_{l,m}(r)$ is in the family of Wendland functions that are shaped like a Gaussian \citep{chernih2014wendland}:
\begin{equation}
\label{wendland}
\phi_{l,m}(r) = \left\{
\begin{array}{cl}
\frac{1}{\Gamma(m) 2^{m-1}} \int_r^1 s(1-s)^l(s^2-r^2)^{m-1}ds &\mbox{for } 0\leq r\leq 1, \\
0  &\mbox{for } r>1.
\end{array}
\right.
\end{equation}

\begin{figure}
\centering
\includegraphics[width=0.75\textwidth]{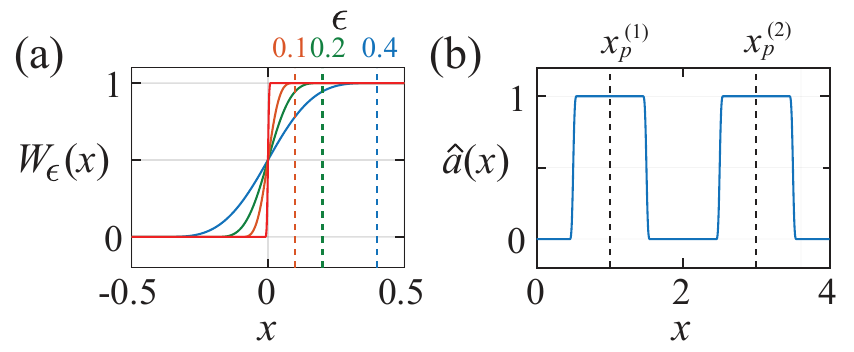}
\caption{Smooth step and indicator functions. (a) Smooth step function $W_\epsilon$ that has a transition interval of $[-\epsilon,\epsilon]$. Four values of $\epsilon = 0.01,0.1,0.2,0.4$ are plotted. (b) Smooth indicator function $\hat{a}$ for locating the region of solid plates. The parameters plotted are $x_p^{(1)} = 1$, $x_p^{(2)} = 3$, $d = 1$, and $\epsilon = 0.05$.}
\label{fig3}
\end{figure}

The integer $m$ controls the smoothness of the Wendland function, and $l = \floor*{m+n/2} + 1$ for spatial dimension $n$. It can be shown that $\phi_{l,m}\in C^{2m}(\mathbb{R}^+)$ and it is compactly supported. Next, we take $m = 1$, $l = 2$ and construct a smooth step function $W_\epsilon(x)$ that transitions from 0 to 1 on $[-\epsilon,\epsilon]$, 
\begin{equation}
    W_\epsilon(x) = \frac{\int_{-\infty}^x \phi_{2,1}(|s|/\epsilon)ds}{ \int_{-\infty}^\infty \phi_{2,1}(|s|/\epsilon)ds}.
\end{equation}
This function is plotted in \cref{fig3}(a) with various $\epsilon$, and it is easy to verify that $W_\epsilon(x) = 0$ when $x < -\epsilon$ and $W_\epsilon(x) = 1$ when $x>\epsilon$. As the transition length $2\epsilon$ becomes smaller, $W_\epsilon$ becomes sharper, approximating a step function. Moreover, $W_\epsilon\in C^3(\mathbb{R})$ when $\epsilon>0$ due to our choice of $m = 1$, achieving our goal of constructing a smooth step function.

We can further evaluate $W_\epsilon(x)$ as
\[
  W_\epsilon(x)=\begin{cases}
               0 \quad \mbox{if } x<-\epsilon,\\
               -\frac{3}{4}\left(\frac{x}{\epsilon}\right)^5+\frac{5}{2}\left(\frac{x}{\epsilon}\right)^4 \mbox{sgn}(x)-\frac{5}{2}\left(\frac{x}{\epsilon}\right)^3+\frac{5}{4}\frac{x}{\epsilon}+\frac{1}{2}\quad  \mbox{if } x\in[-\epsilon,\epsilon],\\
               1 \quad \mbox{if } x>\epsilon.\\
            \end{cases}
\]

We next construct a smooth characteristic function $\hat{\mathbb{1}}_{x\in P}$ for a plate $P$ centered at $x_p$ with length $d$,
\begin{equation}
    \hat{\mathbb{1}}_{x\in P_i}=W_\epsilon\left(x-(x_p-\frac{d}{2})\right) W_\epsilon\left((x_p+\frac{d}{2})-x \right).
\end{equation}

In our numerical examples, we take $\epsilon = 0.05d$ to ensure smoothness. Finally, we can write the Robin boundary conditions in the following form.
\begin{equation}
\label{movingBCsmooth}
  \begin{cases}
    (1-\hat{a})\,\T + \hat{a}\,\T_y = 0\\
    \hat{a}\,u+(1-\hat{a})\,u_y = \hat{g}\quad\quad\mbox{at } y=1,\\
    v = 0
\end{cases}  
\end{equation}
where
\begin{equation}
\label{ahat}
    \hat{a}(x) = \sum_{i}\hat{\mathbb{1}}_{x\in P_i},\,\, \hat{g}(x) =  \sum_{i}u_p^{(i)}\hat{\mathbb{1}}_{x\in P_i}.
\end{equation}
Noticing that $W_\epsilon(0) = 0.5$, it can be easily verified that $\hat{a}(x)\in [0,1]$ as long as $|x_p^{(i)}-x_p^{(j)}|\geq d$ for $i\neq j$. The function $\hat{a}(x)$ for two plates centered at $x_p^{(1)} = 1$, $x_p^{(2)} = 3$ with plate length $d = 1$ is shown in \cref{fig3}(b).

\subsection{Dynamics of the moving plate}
\label{num-boundary-dynamics}
There are two types of forcing on each plate. One is the fluid force due to shear stress, the other is the interaction force when two plates make contact.

For the fluid force, we simply integrate the shear stress by replacing the characteristic function with its smooth version, 
\begin{equation}
    f^{(i)} = -\Pra{}\int_0^\Gamma u_y(x,1,t)\hat{\mathbb{1}}_{x\in P_i}d\,x =  -\Pra{}\Gamma \langle u_y \hat{\mathbb{1}}_{x\in P_i}\rangle.\label{flow-f}
\end{equation}

As we are using an equally spaced periodic grid in $x$, the integration  can be replaced with a numerical average of all grid values of the integrand so $\int_0^\Gamma f(x)\,dx \approx \Gamma \langle f\rangle = (\Gamma/L)\sum_{k=1}^L f_k$, which is spectrally accurate \citep{trefethen2000spectral}.

When solid plates make contact, a contact force between them keeps the plates separated. Inspired by the experiments of \cite{Zhong2005,zhong2007a,zhong2007b}, we set the collision between neighboring plates to be fully elastic, which means their total momentum and kinetic energy are preserved after each collision. Numerically, we approximate the contact force by a function that is both short-ranged and smooth, so the force is zero when the plates are far away but increases rapidly as they get close. 

For plate number $i$, there are two forces for contact from the left neighbor and from the right neighbor,
\begin{align}
  f_{l}^{(i)} &= f_{\max}W_\delta\left(d - |x_p^{(i)}-x_p^{(i-1)}|\right),\label{left-f}\\
  f_{r}^{(i)} &= -f_{\max}W_\delta\left(d - |x_p^{(i+1)}-x_p^{(i)}|\right).\label{right-f}
\end{align}
Here the parameter $\delta$ models an ``interaction length'', and $f_{\max}$ is the maximum interacting force between two plates. With the typical simulation parameters and no fluid forcing, we have verified that this choice of interacting force indeed conserves total momentum and results in a coefficient of restitution $e > 0.99$. At each simulation, $\delta$ and $f_{\max}$ are chosen according to the spatial and temporal resolution, so the ODE and PDE solvers can sufficiently resolve the plate motion and the associated boundary conditions.

\section{Results}
\label{results}
\subsection{One plate dynamics}
\begin{figure}
\centering
\includegraphics[width=0.85\textwidth]{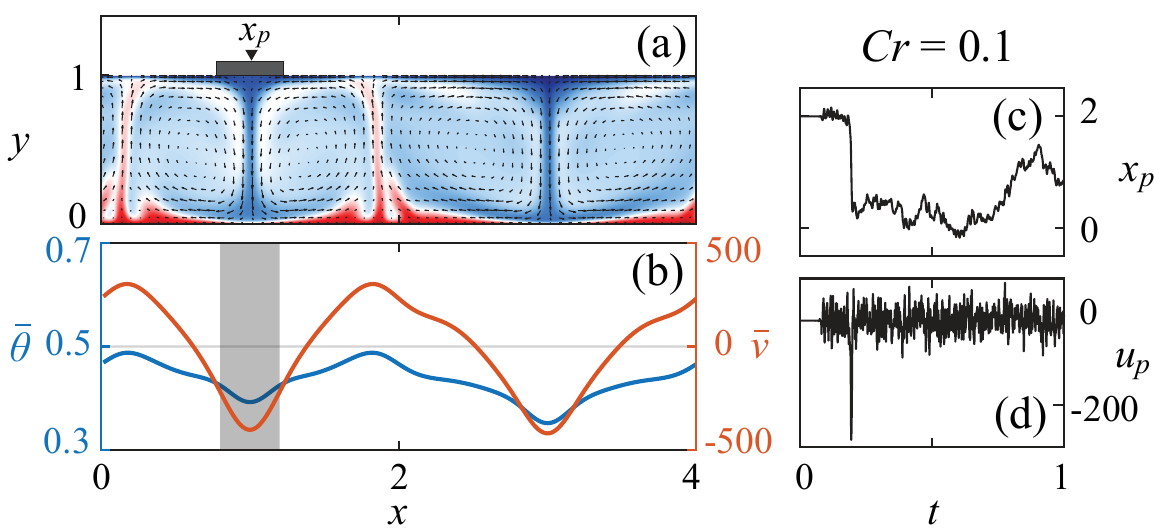}
\caption{ Motion of a small plate ($\Cr{} = 0.1$) is random and bidirectional. (a) A snapshot of flow and temperature fields beneath a plate. The small plate is trapped at a cool converging center. (b) Vertically averaged temperature $\bar\theta$ and vertical velocity $\bar v$ at the same moment of (a). The shaded region indicates the location of the plate. At the converging center, the averaged temperature is low and the flow moves downward. (c)-(d) The displacement $x_p$ and velocity $u_p$ of the plate show behavior of a random walk with jumps. }
\label{fig4}
\end{figure}

In this section, we review the dynamics of a single plate motion. To simplify our study, the Rayleigh number is fixed at $\Ra = 10^6$, the Prandtl number is $\Pra = 7.9$, and the aspect ratio is $\Gamma = 4$. These parameters are similar to previous numerical study \citep{Huang_2024}. For the plate, we set the mass as $\mass = 4d$, so plates with various length $d$ have the same density. For the numerical solver, there are $512$ Fourier modes in the $x$ direction and $129$ Chebyshev nodes in the $y$ direction, and the time step size is $\Delta t = 10^{-6}$. These parameters yield accurate, stable, and resolved numerical solutions.

\begin{figure}
\centering
\includegraphics[width=0.85\textwidth]{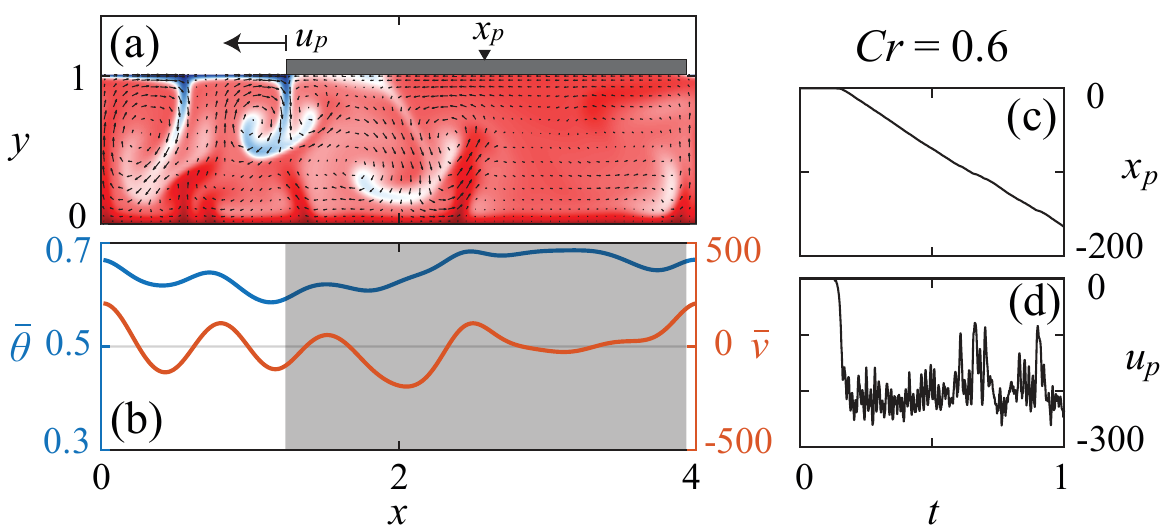}
\caption{ Motion of a large plate ($\Cr{} = 0.6$) is unidirectional. (a)-(b) Flow and temperature fields beneath the plate. (c)-(d) The displacement $x_p$ and velocity $u_p$ of the moving plate, which shows a unidirectional motion with nonzero mean velocity. }
\label{fig5}
\end{figure}

To measure the size of the plate, we define the covering ratio $\Cr{} = d/\Gamma$. Depending on the size of the plate, or $\Cr{}$, the dynamics of the plate motion can be very different. \Cref{fig4} and Supplemental Movie 1 show the dynamics of a small plate with $\Cr{} = 0.1$, and its motion is a continuous random walk shown in \cref{fig4}(c). However in \cref{fig5} and Supplemental Movie 2, a larger plate with $\Cr = 0.6$ shows completely different dynamics: it translates unidirectionally as shown in \cref{fig5}(c).

In \cite{Huang_2024}, the two dynamics are analyzed in detail, and we summarize the key interplay between the plate and the fluid below.

In \cref{fig4}, the small plate tends to be attracted by the converging center of the fluid -- the location where the fluid sinks. This converging center can be clearly seen in \cref{fig4}(a)-(b), located at the minimum of both the vertically averaged temperature $\bar\theta(x,t) = \int_0^1 \theta(x,y,t)\,dy$, and the averaged vertical flow speed $\bar v(x,t) = \int_0^1 v(x,y,t)\,dy$. This means the plate velocity $u_p = \dot{x}_p$ in \cref{fig4}(d) matches the translational velocity of the flow converging center, which has a 0 mean but is subject to noise due to random fluid forcing. In this case, the plate motion is passive and completely driven by the flow structure, and the converging center of the surface flow serves as a stable equilibrium location for the plate.

For a larger plate with $\Cr{} = 0.6$, \cref{fig5} and Supplementary Movie 2 show that the plate motion becomes unidirectional. Increasing plate size clearly changes the flow and temperature distribution in the fluid, as the bulk fluid temperature in \cref{fig5}(a)-(b) is visibly higher than that in \cref{fig4}(a)-(b). This is a clear sign of the thermal blanket effect, as the bigger plate shields the heat from escaping and the effective cooling area at $y=1$ is smaller. In this case, the plate is no longer passive, but creates a thermal blanket that warms the fluid beneath it. Unlike the situation of small plates, a large plate sitting on top of a converging center cannot be stable in the long term, as eventually the temperature beneath the plate will become high enough to turn this converging center into a divergent one. Shown in \cref{fig5}(b), the average temperature $\bar\theta$ is indeed higher below the plate, and the plate sits between the converging and diverging centers. This causes the unidirectional motion of the plate, and as the plate keeps affecting the temperature distribution beneath it, the temperature and flow fields move with the plate as shown in Supplementary Movie 2. The plate displacement $x_p$ and velocity $u_p$ are shown in \cref{fig5}(c)-(d), where $u_p$ has a nonzero mean and is subject to random forcing from the fluid. 

\begin{figure}
\centering
\includegraphics[width=0.8\textwidth]{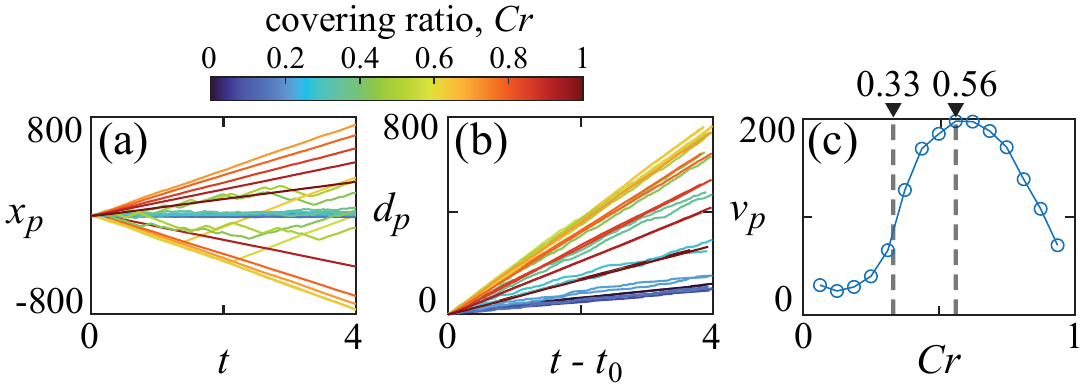}
\caption{Plate displacement and velocity for different covering ratios $\Cr{}$. (a) Sample trajectories of the plate location, where small plates are more affected by noise and large plates have more persistent unidirectional motion. (b) Total travel of the plate reveals its speed, a maximum speed of travel can be seen at around $\Cr{} = 0.6$. (c) Average travel speed has a maximum at $\Cr{} = 0.56$, and unidirectional motions start to appear for plates larger than $\Cr{} = 0.33$.  }
\label{fig6}
\end{figure}

The motions of plates with various $\Cr{}$ are shown in \cref{fig6}. The displacement in \cref{fig6}(a) clearly shows that the small plate has a random motion whose net displacement grows slowly in time. As $\Cr{}$ increases, the plate starts to have more persistent unidirectional motions, however the random fluid forcing can easily reverse the travel direction of the plate, leading to reversals of direction in \cref{fig6}(a). Further increasing $\Cr{}$, the unidirectional motion becomes more persistent, however the velocity (slope of $x_p$) decreases. Defining the total travel of a plate as $d_p(t) = \int_0^t |u_p(t')|\,dt'$, \cref{fig6}(b) shows a peak of the plate traveling speed at around $\Cr{} = 0.6$. To further verify this, we define the average plate speed as $v_p = \lim_{t\to\infty} d_p(t)/t$ in \cref{fig6}(c), and a maximum indeed appears at $\Cr{} = 0.6$.

\begin{figure}
\centering
\includegraphics[width=\textwidth]{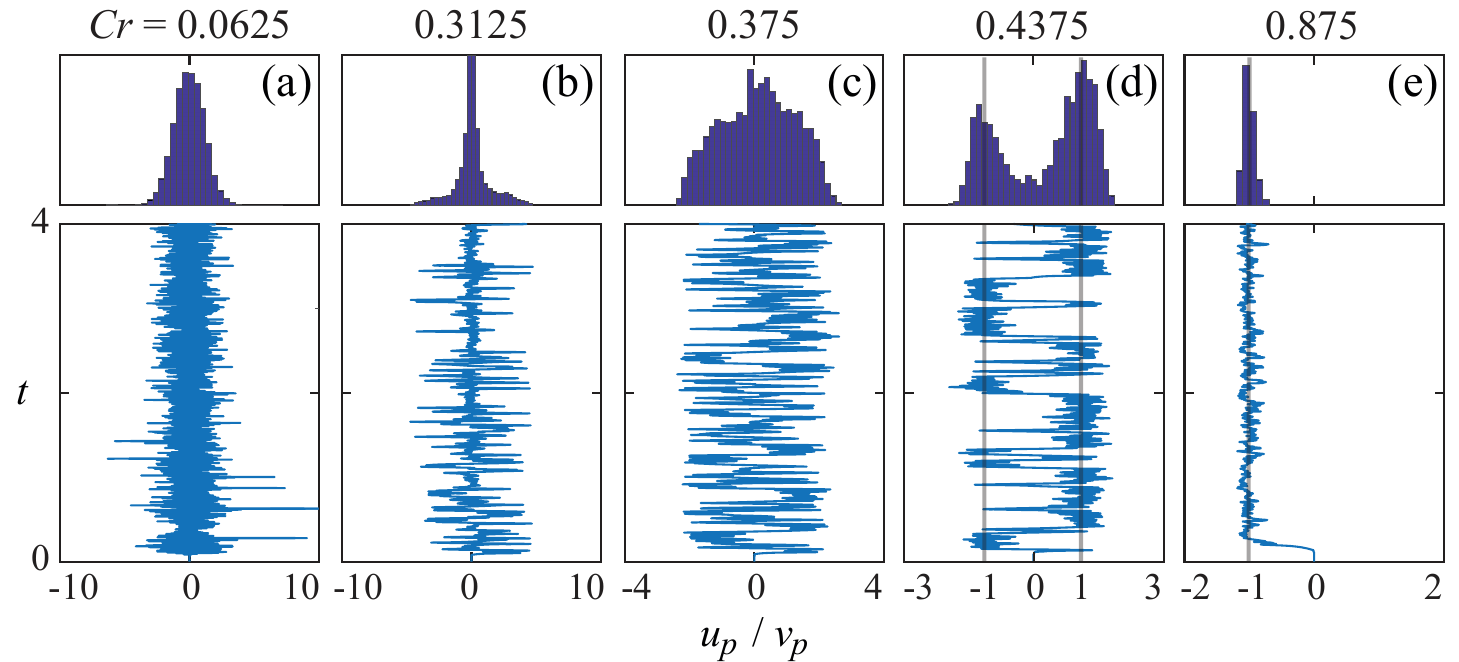}
\caption{ Time series (lower panels) and histogram (upper panels) of the plate velocity $u_p$ at various $\Cr$. The plate velocity is normalized by its average travel speed $v_p$, so $u_p\approx \pm v_p$ suggests a unidirectional translation. (a)-(e) have the covering ratio of $0.0625$, $0.3125$, $0.375$, $0.4375$, $0.875$. The plate motion has a transition from the passive state (a)-(b) to the translating state (d)-(e), and the translation is also more persistent for large $\Cr$ in (e).   }
\label{fig7}
\end{figure}

As the thermal blanket effect strengthens, there is an apparent transition of the plate dynamics. \Cref{fig7} shows the time series (lower panels) and histogram (upper panels) of $u_p$ at various $\Cr$. For small $\Cr$, the histogram of plate velocity resembles a Gaussian distribution, whose zero mean suggests that the net plate displacement would be small. Increasing $\Cr$ beyond $0.3$, the plate dynamics start to transition as \cref{fig7}$(c)$ shows the variance of $u_p$ increases. At $\Cr = 0.4375$ [\cref{fig7}(d)], the two translational states with $u_p = \pm v_p$ emerge, where $u_p$ switches between the two directions due to the noise of fluid forcing. At even higher $\Cr$ [\cref{fig7}(e)], the unidirectional motion is persistent and the reversal becomes rare. The observation here matches the stochastic theory developed in \cite{Huang_2024}, which consists of a simple model that recovers the mechanical and thermal interplay between the plate and the fluid. This stochastic model predicts that there is only a passive state (no net plate motion) for $\Cr < \Cr^*$ where the critical covering ratio is $\Cr^* = 1/3$ for $\Gamma = 4$, and the translational states can only exist for plates with $\Cr > \Cr^*$, indeed matching \cref{fig7}.

\begin{figure}
\centering
\includegraphics[width=0.8\textwidth]{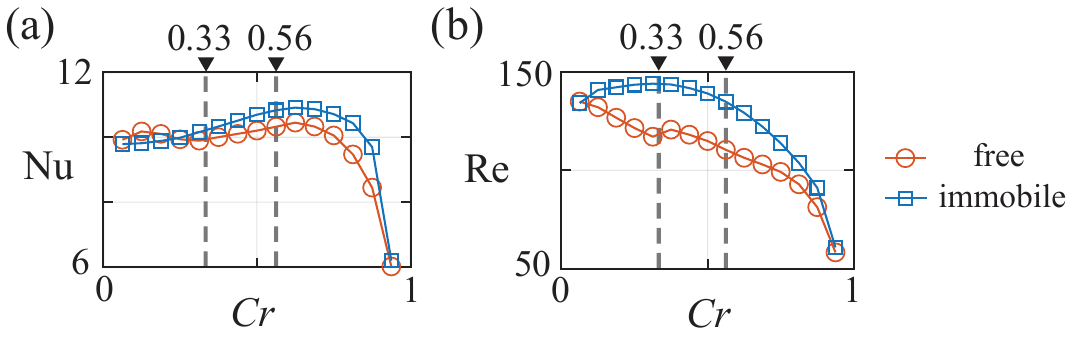}
\caption{ Nusselt and Reynolds numbers for the convecting flow. (a) The Nusselt number is a measure of the vertical heat passing through the flow. (b) Reynolds number as a measure of flow speed. The free data are for the plate freely moving with the flow, and the immobile data are for the plate that is fixed. }
\label{fig8}
\end{figure}

Finally, we investigate how the bulk properties of the flow respond to the moving plate. In \cref{fig8}, we show the Nusselt number $\Nu = - [\Gamma (t_2-t_1)]^{-1}\int_{t_1}^{t_2} dt \int_0^\Gamma \theta_y(x,0,t)\,dx$, and Reynolds number $\ReN = (t_2-t_1)^{-1}\int_{t_1}^{t_2} \left(\max_{(x,y)} |\uu(x,y,t)|\right)\, dt$, where $t_2-t_1$ is the interval for long-time average. The two groups of measurements are for a plate that is free to move by the flow (free), and for a plate that is fixed at a certain location (immobile). By setting the plate free, the Nusselt number changes slightly in \cref{fig8}(a), while the Reynolds number decreases significantly in \cref{fig8}(b). We also note at the critical $\Cr^*$, the flow speed reaches a maximum for the immobile plate shown in \cref{fig8}(b). Moreover, the Nusselt number approaches its maximum at around $\Cr = 0.6$ where the plate translates the fastest as shown in \cref{fig6}(c).  While we do not have a clear theory to explain the observations in \cref{fig8}, we believe the free plate motion certainly modifies the flow and thermal structure of \RBC{}. For example, the fluid shearing drives the plate motion, thus part of the fluid kinetic energy must be converted to the plate kinetic energy. This may explain why a significant decrease of $\ReN{}$ is shown in \cref{fig8}(b).

\subsection{Two-plate interactions}

\begin{figure}
\centering
\includegraphics[width=0.8\textwidth]{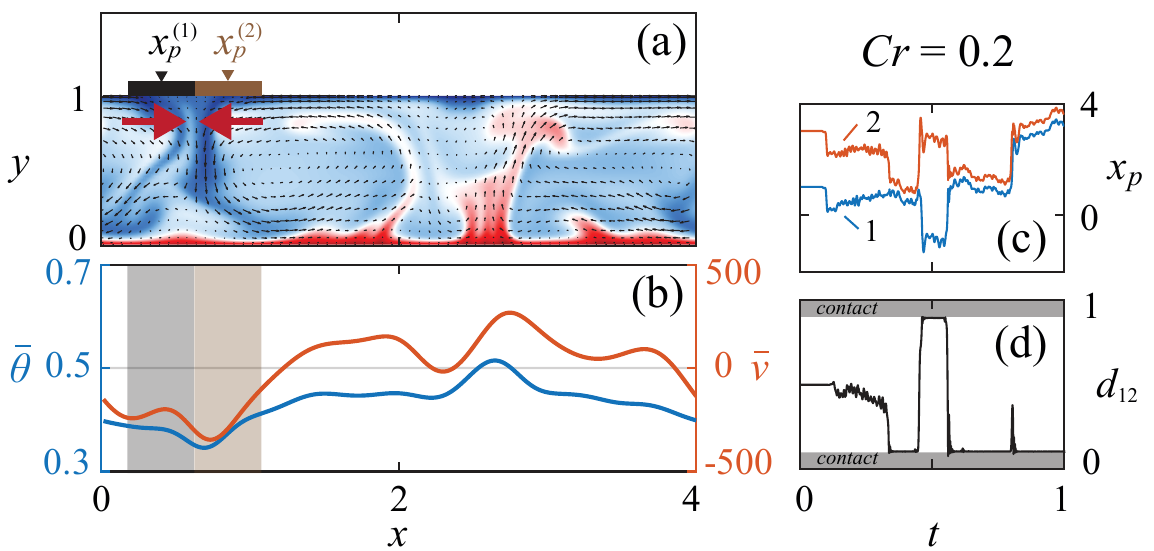}
\caption{ Dynamics of two small plates ($\Cr{}_p = 0.1$ each) forming a supercontinent of $\Cr = 0.2$. (a) Flow and temperature distribution beneath the supercontinent. The surface flow is converging and the formation of the supercontinent is stable. (b) Vertically averaged temperature $\bar\theta$ and vertical velocity $\bar v$ at the same moment of (a), with the region of the two plates shaded. (c) The displacement of plate $x_p^{(1)}$ and $x_p^{(2)}$. (d) The normalized plate distance $d_{12} = [x_p^{(2)}-x_p^{(1)}]/\Gamma$ indicates the two plates tend to stay in contact. The white region (plates separated) and gray region (plates in contact) are separated by $\Cr_p$ and $1-\Cr_p$.}
\label{fig9}
\end{figure}

Adding multiple plates to the convective surface brings interactions between plates and leads to more diverse dynamics. In our numerical simulation, adding a second plate can easily be achieved through the indicator function method outlined in \cref{math,numericalmethods}. In the following numerical experiments, we set $\Ra = 10^6$, $\Pr = 7.9$, $\Gamma = 4$, $m = 4d$ as we did for the single plate case. We additionally set the maximum interaction force $f_{\max{}} = 10^6$ and an interaction range $\delta = \epsilon$ that matches the size of smoothing region of the indicator function in \cref{num-smooth-bd}. These two parameters define the force of interaction between the two plates through \cref{left-f,right-f}, and such interaction conserves both the kinetic energy and momentum of the plates.

\begin{figure}
\centering
\includegraphics[width=0.8\textwidth]{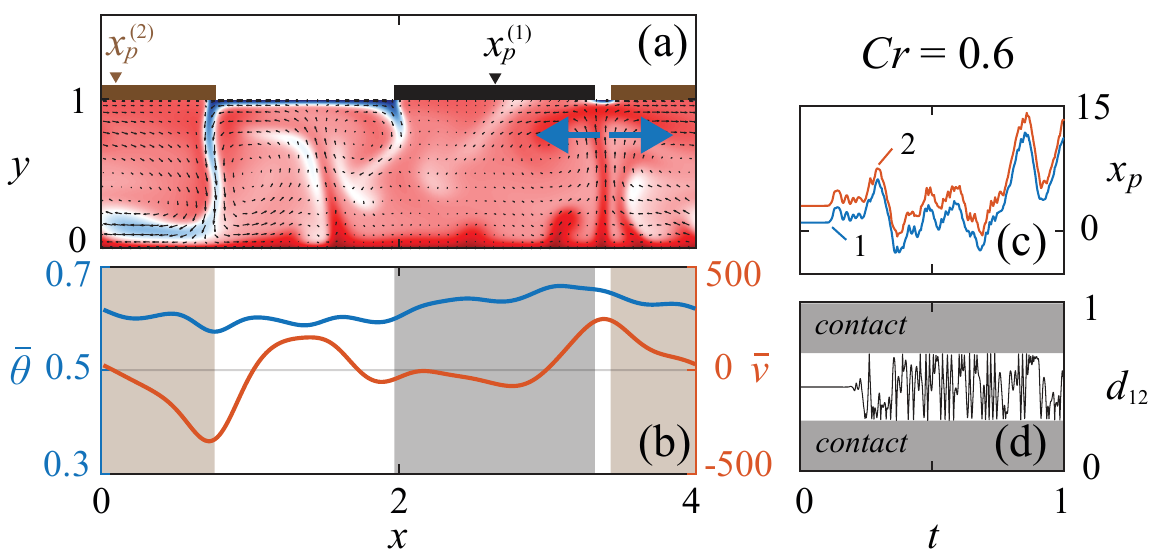}
\caption{Two large plates ($\Cr{}_p = 0.3$ each) cannot form a stable supercontinent of $\Cr =2\Cr{}_p = 0.6$. (a) The warm upwelling fluid creates a diverging surface flow beneath the plates. This diverging surface flow pulls the two plates apart, rendering the supercontinent formation unstable. (b) Vertically averaged temperature and vertical velocity of the fluid beneath the plates shown in (a). (c) Plate displacement $x_p^{(1)}$ and $x_p^{(2)}$. (d) The normalized plate distance $d_{12}$ stays in the white region where the two plates are separated. }
\label{fig10}
\end{figure}

The dynamics of a pair of small plates (\cref{fig9}) and a pair of large plates (\cref{fig10}) are quite different. In \cref{fig9} and Supplementary Movie 3, two small plates with individual covering ratios $\Cr_p = 0.1$ are released on the convective surface. The two plates tend to stay together, generating a ``supercontinent" as shown in \cref{fig9}(a). Further analyzing the flow temperature and surface flow rate in \cref{fig9}(b), we see they are in fact attracted by a converging center of the surface flow, and the surface flow pushes them into each other. The trajectories $x_p^{(1)}$ and $x_p^{(2)}$ of the plates are shown in \cref{fig9}(c), and the normalized plate distance $d_{12} = [x_p^{(2)}-x_p^{(1)}]/\Gamma$ is plotted in \cref{fig9}(d). We clearly see that the two plates prefer to stay in contact, as the normalized distance stays near $\Cr_p$ or $1-\Cr_p$ in \cref{fig9}(d).

The combined covering ratio of these two plates is $\Cr = 2\Cr_p = 0.2$, less than the critical covering ratio of $\Cr^* = 1/3$ we identified earlier. Therefore the thermal blanket effect generated by this supercontinent is not strong enough to heat up the fluid beneath, and the surface flow stays converging and pushing the two plates together. Thus, a supercontinent with combined $\Cr < \Cr^*$ is stable in its formation and exhibits a passive motion.

\Cref{fig10} and Supplementary Movie 4 show the dynamics of two plates with $\Cr = 2\Cr_p = 0.6$. In this case, the fluid beneath the supercontinent is warmed up due to the thermal blanket effect and generates an upwelling flow. The resulting diverging surface flow separates the two plates, leading to an unstable supercontinent formation. \Cref{fig10}(c)-(d) show the plate trajectories and the normalized plate distance, and the two plates are seen to stay separated in \cref{fig10}(d) as their normalized distance is in between $\Cr_p$ and $1-\Cr_p$ and the contact sate is only transient.

In \cref{fig10}, although the covering ratio for each plate $\Cr_p < \Cr^*$, their combined $\Cr = 2\Cr_p>\Cr^*$. The supercontinent, once formed, will become unstable as the warm fluid beneath creates a diverging surface flow that pulls the two plates apart.

\begin{figure}
\centering
\includegraphics[width=0.7\textwidth]{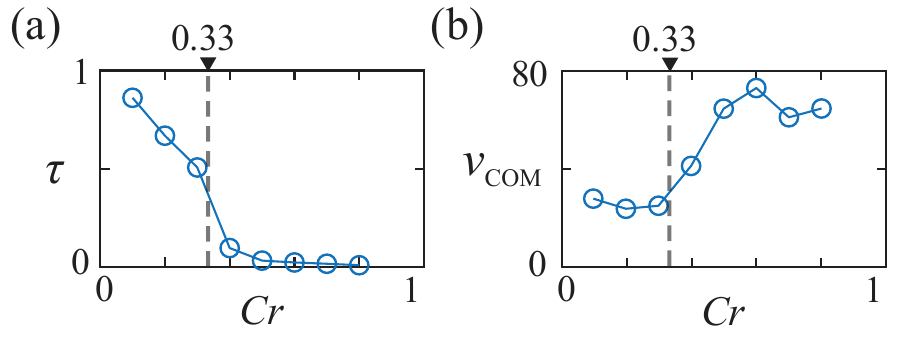}
\caption{ Contact and motion of the plates depend on the covering ratio. (a) The normalized contact time $\tau$ decreases sharply when $\Cr$ increases above $\Cr^* = 1/3$. (b) The plate center of mass velocity $v_{com}$ increases when $\Cr>\Cr^*$, indicating the plates are no longer passive to the flow. }
\label{fig11}
\end{figure}

\Cref{fig11} shows how the plate dynamics depend on the combined $\Cr$. In \cref{fig11}(a), we define a normalized contact time $\tau = t_{c}/T$, where $t_c$ is the amount of time that the two plates are in contact and $T$ is the total simulation time. A sharp decrease of $\tau$ is seen near $\Cr^* = 1/3$, beyond which a persistent formation of supercontinent is unlikely. This is consistent with our analysis earlier, as a supercontinent with $\Cr > \Cr^*$ would induce warm upwelling flows that disintegrate the supercontinent. The center of mass (CoM) velocity of plates also picks up when $\Cr > \Cr^*$ [\cref{fig11}(b)], suggesting the two large plates are no longer passive but instead translate like we have seen in \cref{fig6}(c).

We now see the role of $\Cr^*$ in the formation of supercontinents: It is only possible to have a stable supercontinent if its $\Cr$ is less than $\Cr^*$.

\subsection{Multiple plates}

\begin{figure}
\centering
\includegraphics[width=0.8\textwidth]{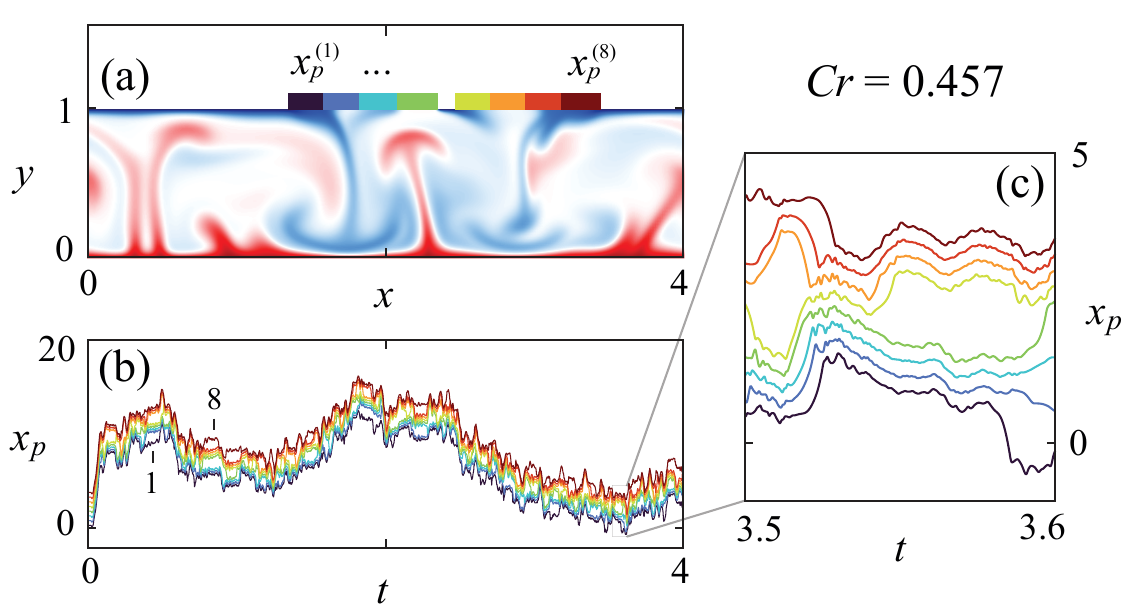}
\caption{ Dynamics of 8 plates ($\Cr{}_p = 0.057$, $\Cr = 8\Cr{}_p = 0.457$) floating on top of the convecting fluid. (a) A snapshot of 8 plates and the convective fluid beneath, the center location of the plates are $(x_p^{(1)},x_p^{(2)},\cdots, x_p^{(8)})$. (b) Trajectories of $(x_p^{(1)},x_p^{(2)},\cdots, x_p^{(8)})$, and plates can be seen forming supercontinents over time. (c) Zoomed-in view of the trajectories in (b) in the time window of $t\in(0.35,0.36)$. }
\label{fig12}
\end{figure}

Further increasing the number of plates, the formation of supercontinents exhibits complex and intriguing dynamics. In \cref{fig12} and Supplementary Movie 4, 8 plates with $\Cr_p = 0.057$ (total $\Cr = 0.457$) are released on top of the convective fluid, where $\Ra = 10^6$, $\Pr = 7.9$, $\Gamma = 4$, $m = 4d$. The plate maximum interaction force $f_{\max{}} = 10^6$ and interaction range $\delta = \epsilon$ are the same as before.

\Cref{fig12}(a) shows a moment when the 8 plates form 2 supercontinents, with each supercontinent covering $4\Cr_p = 0.23$ of the free surface. Each supercontinent thus has a covering ratio less than $\Cr^* = 1/3$ and is thereby stable by our earlier argument. Indeed, the stable formation of a supercontinent of 4 plates can be seen on the plate trajectory $(x_p^{(1)},x_p^{(2)},\cdots, x_p^{(8)})$ shown in \cref{fig12}(b), whose zoomed-in view between $t\in(0.35,0.36)$ is shown in \cref{fig12}(c). Of course, the formation of a supercontinent with a different number of plates is possible, and our theory predicts that they are stable if the number of plates $I<\Cr^*/\Cr_p \approx 6$.

\begin{figure}
\centering
\includegraphics[width=0.8\textwidth]{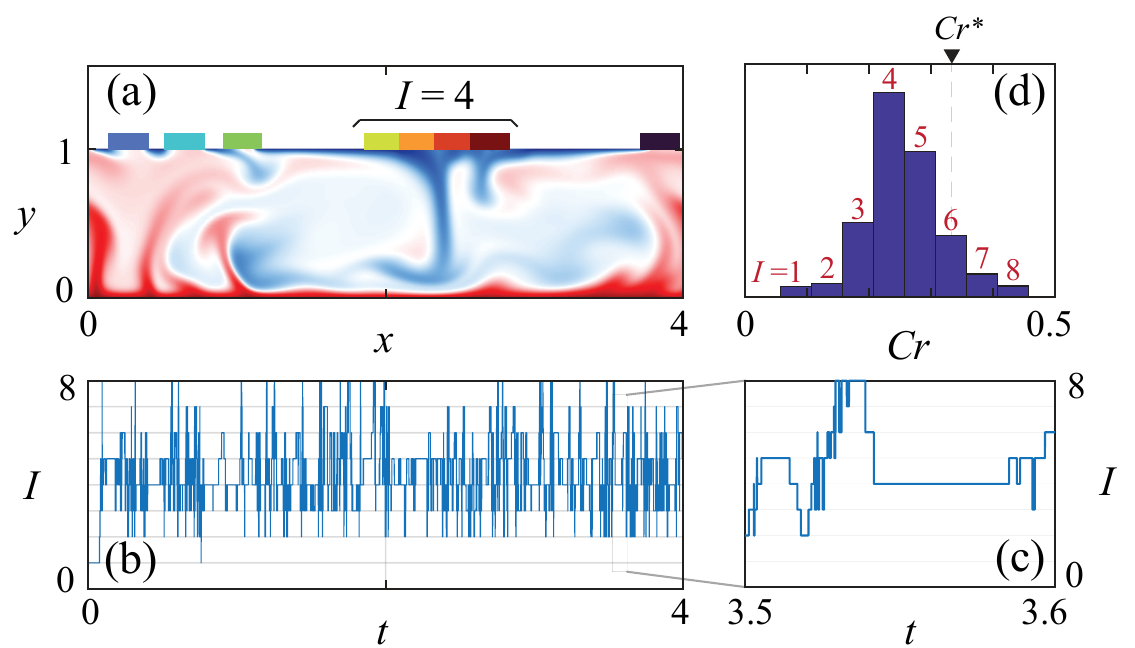}
\caption{ Statistics of the formation number $I(t)$ that is the maximum number of plates in a supercontinent at time $t$. (a) Schematic of $I=4$. (b) Time series of $I(t)$ shows the possibility of forming supercontinents with various sizes. (c) Zoomed-in view of $I(t)$ in (b) during $t\in(0.35,0.36)$. (d) Histogram of $I(t)$ indicates $I=4$ is the most common supercontinent formation, while small and large supercontinents are unlikely to form. The histogram is plotted against the size of supercontinent $\Cr = I\Cr_p$ and the formation number $I$ is labelled on top of each bin.  }
\label{fig13}
\end{figure}

To further analyze the formation of supercontinents, we define the formation number $I(t)$ as the number of plates forming the largest supercontinent at any given time $t$. A schematic of such a formation number is shown in \cref{fig13}(a), where 5 continents are formed and the largest supercontinent has $I = 4$. The formation number can be tracked over time, and \cref{fig13}(b) shows the formation number of the simulation presented in \cref{fig12}, with the zoomed-in view of $I(t)$ in the window of \cref{fig12}(c) shown in \cref{fig13}(c). We note that the formation number can actually take on all integer values between 1 and 8, although many of the formation numbers are transient (such as $I = 1$ and $I = 8$). The most common and persistent formation number we can visually see in \cref{fig13}(c) is $I = 4$, and this is confirmed in the histogram of $I(t)$ shown in \cref{fig13}(d). 

In \cref{fig13}(d), each bin corresponds to the total number of appearances of supercontinents with 
size $I$ (shown on top of each bin), and the horizontal axis shows the corresponding covering ratio. We indeed see that $I = 4$ plates is the most frequent size of a supercontinent. For small supercontinents, they tend to merge and form a larger but stable supercontinent. For large supercontinents, the random fluid forcing can trigger a disintegration and break it into many smaller ones. Especially for $I\geq6$, our theory predicts such formations are not stable, hence supercontinents with $I\geq6$ are rarely formed and (once formed) are unstable.

The results shown in \cref{fig12} and \cref{fig13} are common for simulations with different number ($N_p$) and size ($\Cr_p$) of plates. And we identify $I \Cr_p < \Cr^*$ as a clue for predicting the supercontinental size $I$.

\subsection{Convection with increased $\Ra{}$ and $\Gamma$}
\label{results-large}

In this section, we show examples with Rayleigh number $\Ra{} = 10^7$, Prandtl number $\Pr{} = 7.9$, and domain aspect ratio $\Gamma = 10$. Although these parameters deviate from previous laboratory and numerical investigations, they are actually closer to the conditions of mantle convection \citep{Whitehead2015}.

\begin{figure}
\centering
\includegraphics[width=0.9\textwidth]{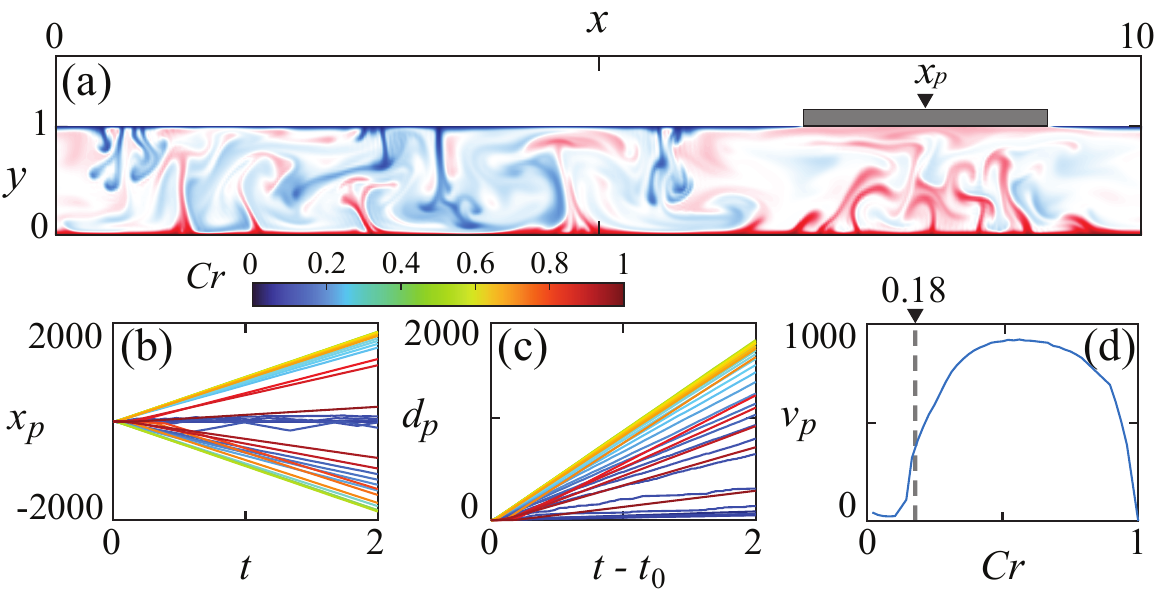}
\caption{Single plate dynamics for large aspect ratio convection, where $\Gamma = 10$, $\Ra{} = 10^7$, and $\Pra{} = 7.9$. (a) Typical convective flow field for a plate with covering ratio 0.2.  (b) Trajectories of the plate location $x_p$, where small plates move passively but large plates translate unidirectionally. (c) Total travel of the plate $d_p$ shows the same trend as in (b). (d) Average travel speed $v_p$ has a sharp increase near $\Cr^* = 0.18$, which is the critical covering ratio for $\Gamma = 10$. Additional simulations of small and large plates can be found in Supplemental Movies 6 and 7. }
\label{fig14}
\end{figure}

We first investigate the single plate dynamics in \cref{fig14}, where the typical flow and temperature fields are shown in \cref{fig14}(a). Much like the observations made in \cref{fig4,fig5,fig6}, \cref{fig14}(b)-(c) show small plates have little motion and are passive to the flow structure, while the large plates translate unidirectionally. Typical simulations for a small plate with $\Cr{} = 0.125$ and large plate with $\Cr{} = 0.417$ are included as Supplemental Movies 6 and 7.

In \cref{fig14}(d), the plate velocity $v_p$ is significantly higher than \cref{fig6}(c) where $\Ra{} = 10^6$. This is consistent with the model introduced in \cite{Huang_2024}, which suggests the equilibrium plate velocity is proportional to the surface flow rate. From the classic scaling $\ReN\sim \Ra{}^{0.5}$ of \RBC{} \citep{Huang2022a}, we infer that the flow rate for $\Ra{} = 10^7$ should be $\sqrt{10}$ times bigger than that of $\Ra{} = 10^6$. The plate velocity in \cref{fig14}(d) is indeed about $3-5$ times higher than the velocity shown in \cref{fig6}(c), thus consistent with our estimation. Therefore the $\Ra{}$ of the convecting fluid directly affects the plate speed, and we note that higher $\Ra{}$ also introduces finer flow structures shown in \cref{fig14}(a) which can potentially modify the strength and distribution of the stochastic fluid forcing.

Another difference between \cref{fig14}(d) and \cref{fig6}(c) is the critical covering ratio $\Cr{}^*$ differentiating the passive and translating states of the plate. In \cref{fig14}(d), a significant increase in $v_p$ appears at around $\Cr{}^* = 0.18$, smaller than $\Cr{}^* = 0.33$ in \cref{fig6}(c) where $\Gamma = 4$. In \cite{Huang_2024}, the critical covering ratio $\Cr^*$ is shown to depend on the aspect ratio as $\Cr^*(\Gamma)\sim\Gamma^{-2/3}$. So $\Cr{}^*(10) = \Cr{}^*(4) (10/4)^{-2/3} = 0.18$, consistent with the data shown in \cref{fig14}(d). Therefore the critical covering ratio $\Cr^*$ decreases with increasing $\Gamma$, while the associated plate length $d^* = \Gamma \Cr^* \sim \Gamma^{1/3}$ increases weakly with $\Gamma$.

\begin{figure}
\centering
\includegraphics[width=0.9\textwidth]{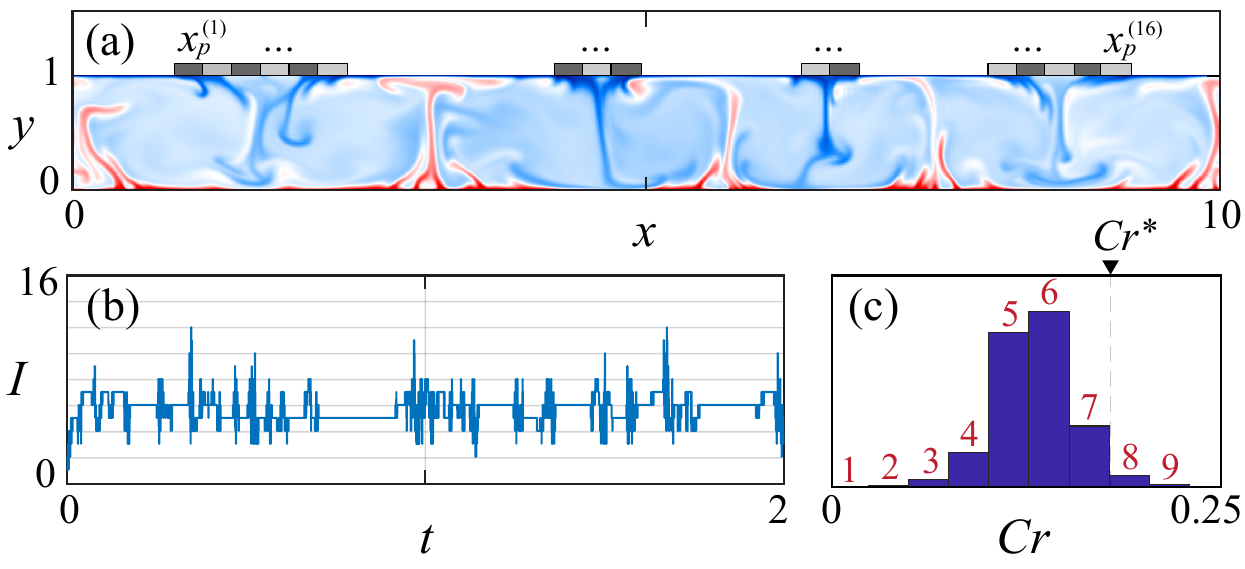}
\caption{Multiple plate dynamics for large aspect ratio convection. There are 16 plates with individual covering ratio $\Cr_p = 0.0234$. The convection parameters are $\Gamma = 10$, $\Ra{} = 10^7$, and $\Pra{} = 7.9$. (a) Typical convective flow field below the 16 moving plates, small groups of supercontinents can be seen.  (b) Formation number $I(t)$ indicating the size of the largest supercontinent at time $t$. (c) Histogram of the formation number $I$ showing $I = 6$ is the most probable formation of supercontinents, and the formation of superconinents with covering ratio above critical is rare.  Movie associated with this simulation can be found as Supplemental Movie 8. }
\label{fig15}
\end{figure}

We finally look at the formation of supercontinents at high $\Ra{}$ and $\Gamma$. \Cref{fig15}(a) shows a typical moment of supercontinent formation, where continents formed by 6, 3, 2, and 5 plates can been seen. In \cref{fig15}, each plate has a covering ratio $\Cr_p = 0.0234$, and the movie of this simulation is included as Supplemental Movie 8. 

In \cref{fig15}(a), we can see the largest supercontinent formation still sits on a converging center of surface flow, thus making such a formation stable. Like we have done in the previous section, we define the formation number $I$ as the size of the largest supercontinent at a given time, so $I = 6$ for the moment shown in \cref{fig15}(a). The time series of this formation number is shown in \cref{fig15}(b), where a formation size of around 5-6 plates is common but both the large and small formations are rare. The histogram of $I(t)$ in \cref{fig15}(c) also shows the most probable supercontinent formation has $I = 6$, whose total covering ratio $\Cr{} = I\Cr{}_p =  0.14$ is smaller than the critical covering ratio $\Cr{}^* = 0.18$ we identified earlier. In fact, \cref{fig15}(c) shows the formation of $I = 7$ is not rare either, and this is consistent with our estimation that the largest stable supercontinent can have a size up to $\Cr{}^*/\Cr{}_p = 7.7$.

In the discussions above, we see once again that the condition $I\Cr{}_p < \Cr{}^*$ can serve as an indicator of stable supercontinent formation.

\section{Discussion}
\label{discussion}

Through our numerical investigations, we clearly see the covering ratio as the main factor affecting the thermal blanket effect, which determines the plate dynamics. This is especially apparent during the formation of supercontinents-- the continent covering ratio cannot exceed $\Cr^*$, as the strong thermal blanket effect will induce a diverging surface flow that pulls the formed supercontinent apart.

As our current study is inspired by laboratory experiments, we certainly look forward to future experimental investigations of the interaction between multiple plates. Besides the geometry presented in \cref{fig1,fig2}, the broader investigation of fluid-structure interactions in thermal convection also includes adding fixed obstacles to the convective domain \citep{Bao_Chen_Liu_She_Zhang_Zhou_2015,Li_Chen_Xi_2024}, modifying the convective boundary conditions \citep{zhang2020controlling,Huang2022a}, and allowing for moving boundaries that are driven by the fluid forcing \citep{Mercier2014,wang2023persistent} or phase change \citep{mac2021stable,mac2022morphological,Weady2022}. Many of these works are certainly geophysically inspired, addressing long-standing mysteries such as the super rotation of Earth's inner core \citep{wang2023persistent,yang2023multidecadal}. 

In order to more closely simulate the geophysical plate tectonics, there are several future directions to improve our model. The first one is to investigate the plate tectonics in a 3-dimensional geometry, including the mantle-like spherical shell and the rectangular cuboidal fluid domain that is periodic in two horizontal directions. Although our current numerical scheme is 2-dimensional, it can be easily extended to a higher dimension: For the spherical shell geometry, we can adopt a Fourier-Chebyshev-Chebyshev method that solves the heat and flow equations in spherical coordinates; For the periodic cuboidal domain, another periodic horizontal direction can be accommodated by the Fourier method. We are currently working on the 3-dimensional study of plate tectonics, and we note that pioneer works including \cite{lowman1999thermal,mao2019dynamics,mao2021insulating} have investigated the fluid-structure interactions and plate interactions in these settings. They also used the geophysical parameters of the mantle, most notably a Prandtl number around $10^{23}$ \citep{meyers1987encyclopedia}. One modification to accommodate this high Prandtl number is to take an asymptotic limit of $\Pr\to\infty$, so the Navier-Stokes \cref{NS} becomes a time-independent Stokes equation \citep{mao2021insulating}.  This Stokes equation can be easily handled by the numerical method in \cref{numericalmethods}, however we choose to keep the fluid inertia so our results can be validated by future laboratory experiments, where the working fluid (water) has  $\Pr = 1-10$.

In our study, we have also set the bottom boundary condition to be no-slip, which is consistent with the experiments. However this is different from past geophysical models of the mantle, where the bottom boundary is typically assumed as stress-free at the liquid-liquid interface between the mantle and the outer core of Earth. In the future, we plan to adopt the stress-free bottom condition as it is more consistent with the geophysical settings. Descending deeper into Earth, the outer core meets the solid inner core that is known to rotate. Together with the recent discovery that this rotation is not unidirectional \citep{yang2023multidecadal}, a fluid-structure interaction problem thus arises: Does the outer-core convection provide enough shear stress to rotate the inner core? This possibility has inspired some recent experimental developments \citep{wang2023persistent}, and we wish to further engage with this fluid-structure-interaction problem in the future.

The interaction between continental plates is also more complicated in geophysical plate tectonics, as the converging continental plates deform the contact region and form the tallest mountains of Earth. It is still an ongoing quest to understand the consequences of converging and diverging continents, with many recent works focusing on the deforming contact region and addressing its influence on the mantle flow beneath \citep{rozel2017continental,whitehead2023convection}. Inspired by the geophysics, one modification to our current model could be changing the way neighboring plates interact. This can be achieved through changing the coefficient of restitution to 0 during each collision, which better captures the collision between continental plates during a long time scale. Adding an attractive force between plates can also reveal interesting dynamics, as the diverging surface flow has to be strong enough to pull the plates apart. Thus, a larger formation of translating supercontinent may stay stable in this case, and the contact state of plates would also become more persistent. Looking forward, more sophisticated models are required to capture different plate interactions and resulting plate dynamics, and they may reveal new insights into the physics and dynamics of the formation of supercontinents.

To conclude, we consider a toy model to predict the size of the largest continental plate. The aspect ratio of Earth's mantle is roughly $\Gamma_E = 10$, so the critical covering ratio there is $\Cr^*_E  \approx 0.18$ as we have estimated in \cref{results-large}. This gives an estimated dimension of the largest stable continental plate of Earth, $L \approx 2\pi R_E \Cr^*_E$, where $R_E = 6400$ km is the radius of Earth. So the plate area is approximately $L^2 \approx 6\times10^7$ km$^2$, slightly underestimating the largest continental plate, the North American Plate, whose area is $7.59\times 10^7$ km$^2$.

\section*{Supplementary Materials}
Supplementary movies are available at \url{https://math.nyu.edu/~jinzi/research/convectivePlate-np/Movie}.

\section*{Declaration of Interests}
The author reports no conflict of interest.

\section*{Acknowledgement}
J.M.H. thanks Jin-Qiang Zhong, Junjun Chu, and Jun Zhang for useful discussions. J.M.H. acknowledges support from the National Natural Science Foundation of China (12272237, 92252204).

\section*{Appendix A: Helmholtz solver for Dirichlet problems}
\label{helmholtz-dirichlet}
We outline the numerical solver for the following Helmholtz problem, 
\begin{align}
    \Lap u - \sigma u &= h(x,y),\\
    u(x,0) &= g_0(x),\\
    u(x,1) &= g_1(x).
\end{align}

As the domain is periodic in $x\in(0,\Gamma]$, we can discretize $x$ as $L$ equally-spaced nodes so $x_l = l\updel x$ where $\updel x = \Gamma/L$ and $l = 1,2,\dots L$. We further require $L$ to be odd. We can now approximate the solution $u(x,y)$ as a truncated Fourier series in $x$,
\begin{equation}
    \label{fourieru}
    u(x_l,y) = \sum_{k=-\frac{L-1}{2}}^{\frac{L-1}{2}} \hat{u}_k(y)\exp{(2\pi \im k l/L)},
\end{equation}
where $\im$ is the imaginary unit. The Fourier coefficients $\hat{u}_k$ then satisfy the ODE,
\begin{align}
    \frac{d^2\hat{u}_k}{dy^2} - \sigma_k \hat{u}_k &= \hat{h}_k(y),\label{ode1}\\
    \hat{u}_k(0) &= \hat{g}_{0,k},\label{ode2}\\
    \hat{u}_k(1) &= \hat{g}_{1,k},\label{ode3}
\end{align}
where
\begin{align}
    \sigma_k &= \frac{4\pi^2k^2}{\Gamma^2} + \sigma,\\
    \hat{h}_k(y) &= \frac{1}{L}\sum_{l = 1}^{L} h(x_l,y) \exp{(-2\pi \im k l/L)},\label{hhat}\\
    \hat{g}_{0,k}&= \frac{1}{L}\sum_{l = 1}^{L} g_0(x_l) \exp{(-2\pi \im k l/L)},\label{g0hat}\\
    \hat{g}_{1,k} &=\frac{1}{L}\sum_{l = 1}^{L} g_1(x_l) \exp{(-2\pi \im k l/L)}.\label{g1hat}
\end{align}
The computation of \cref{hhat,g0hat,g1hat} can be done efficiently with the help of the Fast Fourier Transformation (FFT) algorithm.

Now all that is left is a set of ordinary differential \cref{ode1,ode2,ode3}, which can be solved with methods like finite-differences. We instead use the Chebyshev method and discretize $y\in[0,1]$ to $M+1$ Chebyshev nodes, so $y_m = [1+\cos(m\pi/M)]/2$ with $m = 0,1,2,\dots M$. An advantage of using the Chebyshev method is that the unevenly distributed Chebyshev nodes have a higher resolution near the boundary $y=0$ and $y=1$, therefore resolving the boundary layer structures. The differentiation operator $\frac{d}{dy}$ can be approximated by a discrete operator $\mathbf{D}$ \citep{trefethen2000spectral,peyret2002spectral}, whose elements are
\begin{align}
    D_{j,k} &= 2\frac{c_j}{c_k}\frac{(-1)^{j+k}}{\cos(j\pi/M)-\cos(k\pi/M)} &0\leq j,\,k\leq M,\,\, j\neq k,\\
    D_{j,j} &= -\frac{\cos(j\pi/M)}{1-\cos^2(j\pi/M)}, &1\leq j\leq M-1,\\
    D_{0,0} &= -D_{M,M} = \frac{2M^2+1}{3}.&
\end{align}
Here $c_0=c_M = 2$, $c_j = 1$ for $1\leq j\leq M-1$.

The discrete operation on the LHS of \cref{ode1} can therefore be written as 
\begin{equation}
\label{Adef}
    \mathbf{A} = \mathbf{D}^2-\sigma_k\mathbf{I},
\end{equation}
where $\mathbf{I}$ is a $M+1$ by $M+1$ identity matrix.

Noticing that the Chebyshev nodes are $\mathbf{y} = [1, y_1, y_2, \dots,y_{M-1},0]^\top$,  we can write the discrete solution to \cref{ode1,ode2,ode3} at these locations as a column vector,
\begin{equation}
    \mathbf{U} = [\hat{g}_{1,k}, \hat{u}_k(y_1), \hat{u}_k(y_2),\dots, \hat{u}_k(y_{M-1}), \hat{g}_{0,k}]^\top = [\hat{g}_{1,k}, \Tilde{\mathbf{U}}, \hat{g}_{0,k}]^\top.
\end{equation}
The interior solution $\Tilde{\mathbf{U}}$ therefore satisfies
\begin{equation}
\label{linearsys}
    \Tilde{\mathbf{A}}\Tilde{\mathbf{U}} = \Tilde{\mathbf{H}},
\end{equation}
where the $(M-1)\times(M-1)$ matrix $\Tilde{\mathbf{A}}$ is the interior of $ \mathbf{A}$ (by removing its first and last rows and columns), and 
  \begin{align}
    \Tilde{\mathbf{H}} &= \begin{bmatrix}
           \hat{h}_k(y_1)-\hat{g}_{1,k}A_{1,1} - \hat{g}_{0,k}A_{1,M} \\
           \hat{h}_k(y_2)-\hat{g}_{1,k}A_{2,1} - \hat{g}_{0,k}A_{2,M} \\
           \vdots \\
           \hat{h}_k(y_{M-1})-\hat{g}_{1,k}A_{M-1,1} - \hat{g}_{0,k}A_{M-1,M}
         \end{bmatrix}.
  \end{align}

\Cref{linearsys} is invertible, and the operator $\Tilde{\mathbf{A}}$ does not change during time stepping while all the boundary information $g$ and forcing $h$ are contained in $\Tilde{\mathbf{H}}$. This allows us to compute the LU decomposition of $\Tilde{\mathbf{A}}$ at the beginning so $\Tilde{\mathbf{A}} \Tilde{\mathbf{U}} =  \Tilde{\mathbf{H}}$ can be efficiently inverted through backward and forward substitutions during time stepping. With $\Tilde{\mathbf{U}}$ and $\mathbf{U}$ solved, the Fourier coefficients $\hat{u}_k$ can then be inserted into \cref{fourieru} and the solution $u(x,y)$ is therefore obtained.

\section*{Appendix B: Helmholtz solver for Robin boundary conditions}
Next we consider the Helmholtz solver for equations like \cref{Tdisc}, where inhomogeneous Robin boundary conditions like \cref{movingBC} are applied. In a general form, consider 
\begin{align}
    \Lap u - \sigma u &= h,\label{robinode}\\
    a_0(x)u(x,0)+b_0(x) &u_y(x,0) = g_0(x),\label{robin1}\\
    a_1(x)u(x,1)+b_1(x) &u_y(x,1) = g_1(x).\label{robin2}
\end{align}

The idea of solving these equations is to use the influence matrix method \citep{peyret2002spectral}: A PDE with inhomogeneous Robin boundary conditions can be converted into a series of PDEs with homogeneous Dirichlet boundary conditions, which can be solved by the method detailed in \cref{helmholtz-dirichlet}. 

We first separate the solution into several subproblems, so 
\begin{equation}
    u(x,y) = \Tilde{u}(x,y) + \sum_{l = 1}^{2L} \xi_l \Bar{u}_l(x,y),
\end{equation}
where $\xi_l$ are unknown coefficients to be determined later and 
\begin{align}
    \Lap \Tilde{u} - \sigma \Tilde{u} &= h,\label{tilde1}\\
    \Tilde{u}(x,0) = \Tilde{u}(x,1) &= 0,\label{tilde2}\\[6pt]
    \Lap \Bar{u}_l - \sigma \Bar{u}_l &= 0,\label{bar1}\\
    \Bar{u}_l|_{\eta_m} = \delta_{lm} \mbox{ for all } &\eta_m\in \partial\Omega.\label{bar2}
\end{align}
Here $\eta_m$ represents the $m$th node on the boundary, so there are $2L$ of them, and $\delta_{lm}$ is the Kronecker delta function. Now, the boundary conditions \cref{robin1,robin2} indicate
\begin{equation}
\label{infMsys}
    \left[a_i(x) \left(\sum_{l=1}^{2L} \xi_l\Bar{u}_l\right) +  b_i(x) \left(\pd{\Tilde{u}}{y}+\sum_{l=1}^{2L} \xi_l\pd{\Bar{u}_l}{y}\right)-g_i(x) \right]\biggr\rvert_{\eta_m} = 0 \mbox{  for } \eta_m\in \partial\Omega_i.
\end{equation}
Here $i\in\{0,1\}$ indicates the bottom or top boundary. As \cref{infMsys} holds on all the boundary nodes $\eta_m$, it provides $2L$ equations for $2L$ unknowns $\xi_l$, and such a linear system is invertible. With all $\xi_l$ solved, the solution of \cref{robinode,robin1,robin2} can be recovered as $u = \Tilde{u} + \sum_{l = 1}^{2L} \xi_l \Bar{u}_l$.

In fact, $\xi_m$ is exactly the Dirichlet data for $u$ at boundary node $\eta_m$, therefore the solution of \cref{robinode,robin1,robin2} is the same as the solution of
\begin{align}
    \Lap u - \sigma u &= h,\label{dirichletode}\\
    u|_{\eta_m} &= \xi_m \mbox{  for } \eta_m\in \partial\Omega. \label{newdirichlet}
\end{align}

This method can directly solve the heat \cref{Tdisc} with boundary conditions in \cref{bottom_cond,movingBC-psi}, by assigning $a_0 = 1$, $b_0 = 0$, $g_0 = 1$ at the bottom and $a_1 = 1-a$, $b_1 = a$, $g_1 = 0$ at the top. Usually, the solutions of \cref{bar1,bar2} are obtained and saved at the beginning. At each time step, \cref{tilde1,tilde2} are solved and the location of moving plates will determine $a_i(x)$, $b_i(x)$, and $g_i(x)$, so \cref{infMsys} can be inverted to provide $\xi_m$ which can be used as boundary data in \cref{dirichletode,newdirichlet}, and the solution $\T_n$ can therefore be determined.

\section*{Appendix C: Flow solver with Robin boundary conditions}

At each time step, the following flow problem must be solved,
\begin{align}
    \Lap \omega -\sigma\omega  &= f,\\
    -\Lap \psi &= \omega,\\
    \psi = \psi_y &= 0 \quad\quad\mbox{at } y=0, \label{neumann}\\
    \psi = 0, \quad a\,\psi_y+(1-a)\,\psi_{yy} &= g\quad\quad\mbox{at } y=1. \label{robin3}
\end{align}

Here we have dropped all the subscripts. We can also solve these equations with the influence matrix method. Now we want to convert the Neumann boundary condition in \cref{neumann} and Robin boundary condition in \cref{robin3} to a  Dirichlet boundary condition for vorticity $\omega$. We decompose $\omega$ and $\psi$ as,
\begin{align}
    \omega(x,y) = \Tilde{\omega}(x,y) + \sum_{l = 1}^{2L} \xi_l \Bar{\omega}_l(x,y), \label{decomp1}\\
    \psi(x,y) = \Tilde{\psi}(x,y) + \sum_{l = 1}^{2L} \xi_l \Bar{\psi}_l(x,y).\label{decomp2}
\end{align}
The associated subproblems are
\begin{align}
    \Lap \Tilde{\omega} - \sigma \Tilde{\omega} &= f,\,\, -\Lap \Tilde{\psi} = \Tilde{\omega}\label{flow-tilde1}\\
    \Tilde{\omega}(x,0) = \Tilde{\omega}(x,1) &= \Tilde{\psi}(x,0) = \Tilde{\psi}(x,1) = 0\label{flow-tilde2}\\[6pt]
    \Lap \Bar{\omega}_l - \sigma \Bar{\omega}_l &= 0,\,\, -\Lap \Bar{\psi}_l = \Bar{\omega}_l \label{flow-bar1}\\
    \Bar{\psi}_l(x,1) = 0,\,\, \Bar{\omega}_l|_{\eta_m} &= \delta_{lm} \mbox{ for all } \eta_m\in \partial\Omega.\label{flow-bar2}
\end{align}

The Neumann boundary condition in \cref{neumann} and the Robin boundary condition in \cref{robin3} can now be enforced as
\begin{align}
    \left(\pd{\Tilde{\psi}}{y}+\sum_{l=1}^{2L} \xi_l\pd{\Bar{\psi}_l}{y}\right) \biggr\rvert_{\eta_m} = 0 &\mbox{  for } \eta_m\in \partial\Omega_0,\label{infMomega1}\\
    \left[a \left(\pd{\Tilde{\psi}}{y}+\sum_{l=1}^{2L} \xi_l\pd{\Bar{\psi}_l}{y}\right) +  (1-a) \left(\pd{^2\Tilde{\psi}}{y^2}+\sum_{l=1}^{2L} \xi_l\pd{^2\Bar{\psi}_l}{y^2}\right) -g \right]\biggr\rvert_{\eta_m} = 0 &\mbox{  for } \eta_m\in \partial\Omega_1.\label{infMomega2}
\end{align}
\Cref{infMomega1,infMomega2} are again $2L$ equations for $2L$ unknowns $\xi_l$ and the linear system is invertible. \Cref{decomp1,decomp2} can then be summed and the solutions are obtained. In fact,  $\xi_l$ is exactly the boundary data for $\omega$ at boundary node $\eta_l$, so we can instead solve
\begin{align}
    \Lap \omega -\sigma\omega  &= f,\\
    -\Lap \psi &= \omega,\label{omega-ode1}\\
    \psi|_{\eta_m} = 0,\,\, \omega|_{\eta_m} = \xi_m &\mbox{  for } \eta_m\in \partial\Omega. \label{omega-ode2}
\end{align}

\section*{Appendix D: Summary of numerical methods}

There are two stages during our numerical simulation and below we list some of the key steps during each stage.

\vspace{1em} Preprocessing stage:
\vspace{5pt}
\begin{enumerate}
    \item The inverse $\Tilde{\mathbf{A}}^{-1}$ (or the LU decomposition of $\Tilde{\mathbf{A}}$) in \cref{linearsys} is prepared for operators in \cref{omegadisc,Tdisc,psidisc}, by taking $\sigma = 0, \sigma_1, \sigma_2$ in \cref{Adef};
    \item Subproblems \cref{bar1,bar2,flow-bar1,flow-bar2} are solved, and the solutions $\bar{\theta}_l$, $\bar{\omega}_l$, and $\bar{\psi}_l$ for $l = 1,2,\dots,2L$ are saved. 
\end{enumerate}

\vspace{2em} At the $n$th step of the time-stepping stage: 
\vspace{5pt}
\begin{enumerate}
    \item The fluid and interaction forces on each plate are computed according to \cref{flow-f,left-f,right-f};
    \item The location and velocity of each plate are evolved by \cref{loc,spd}, and the indicator function $\hat{a}$ is prepared by \cref{ahat};
    \item The forcing term $h_n$ in \cref{Tdisc} is prepared according to \cref{hn}, and  $\Tilde{\theta}_l$ is solved from \cref{tilde1,tilde2};
    \item \Cref{infMsys} is inverted so the Dirichlet boundary data for $\theta_n$ is known, \cref{dirichletode,newdirichlet} are then solved for $\theta_n$;
    \item The temperature $\theta_n$ is used to prepare $f_n$ according to \cref{fn}, so $\Tilde{\omega}_l$, $\Tilde{\psi}_l$ can be solved from \cref{flow-tilde1,flow-tilde2};
    \item \Cref{infMomega1,infMomega2} are inverted so the Dirichlet data for $\omega_n$ is known, \cref{omega-ode1,omega-ode2} are finally solved to provide $\omega_n$ and $\psi_n$.
\end{enumerate}

\vspace{2em}  For simulations with $\Gamma = 4$, we use $L = 256$ Fourier nodes and $M+1 = 65$ Chebyshev nodes; For simulations with higher aspect ratio $\Gamma = 10$, $L = 512$ Fourier nodes and $M+1 = 65$ Chebyshev nodes are used instead. During time stepping, we typically set $\updel t = \tau_0 =  5\times 10^{-4} \,\Ra^{-1/2}$ considering that the flow speed $|\uu|\sim \sqrt{\Ra}$. 

These parameters are tested to yield spatially and temporally resolved results. In \cref{tab:kd},  time-averaged values of $\Nu{}$, $\ReN{}$, and $v_p$ (plate speed) are measured at the dynamical equilibrium state of convection with $\Ra{} = 10^7$, $\Pr{} = 7.9$, and $\Gamma = 10$, where a floating plate with $\Cr{} = 1/2$ is free to move on top. The convergence of these values shows our choice of spatial and temporal resolution is sufficient to resolve the dynamics of the flow and the plate.

\begin{table}
  \begin{center}
\def~{\hphantom{0}}
  \begin{tabular}{ccccccc} 
      &$L$  & $M$   &   $\Delta t$ & $\Nu{}$ & $\ReN{}$ & $v_p$ \\[3pt]
      & 256   & 40 & $2\tau_0$ & 13.4 & 492 & 963 \\
      & 256   & 40 & $\tau_0 $ & 13.3 & 492 & 964 \\
      & 256   & 40 & $0.5 \tau_0$ & 13.4 & 493 & 965 \\
      & 512   & 64 & $2\tau_0$ & 12.7 & 489 & 928 \\
     \textasteriskcentered{} & 512   & 64 & $\tau_0 $ & 12.8 & 491 & 925 \\
      & 512   & 64 & $0.5 \tau_0$ & 12.8 & 493 & 920 \\
      & 1024   & 128 & $2\tau_0$ & 12.8  & 493  & 918  \\
      & 1024   & 128 & $\tau_0 $ & 12.8  & 490  & 915 \\
      & 1024   & 128 & $0.5 \tau_0$ & 12.8  & 490  &  915\\
  \end{tabular}
  \caption{Time-averaged dynamical quantities at different spatial and temporal resolutions for plate tectonics with $\Ra{} = 10^7$, $\Pr{} = 7.9$, $\Gamma = 10$, and $\Cr{} = 1/2$. Here $L$ is the number of Fourier modes, $M+1$ is the number of Chebyshev nodes, and $\Delta t$ is the time step size. The asterisked parameters are used in the DNS of \cref{fig14,fig15}, where $\Delta t = \tau_0 = 5\times 10^{-4} \,\Ra^{-1/2}$.}
  \label{tab:kd}
  \end{center}

\end{table}


\bibliographystyle{tex-library/jfm}
\bibliography{refs}
\end{document}